\documentclass[lettersize]{IEEEtran} %onecolumn

\usepackage[utf8]{inputenc}
\usepackage{color,xcolor}
\usepackage{epsfig}
\usepackage{amsmath,amsfonts,amsthm,amssymb}
\usepackage{graphics,graphicx,pgfplots,tikz}%,subfigure}
\usepackage{float}
\usepackage{algorithm,algorithmic}
\usepackage{svg}
\usepackage{epstopdf}
\usepackage{soul,csquotes,ulem}
\usepackage[hidelinks]{hyperref}
\usepackage[font=footnotesize]{subcaption}
\usepackage[font=footnotesize]{caption}

\usepackage[subtle]{savetrees}
\usepackage{flushend}
\usepackage[skip=4pt]{caption}
\usepackage{titlesec}

\titlespacing*{\section}{0pt}{5pt}{5pt} % Adjusts \section
\titlespacing*{\subsection}{0pt}{5pt}{5pt} % Adjusts \section
\usepackage{indentfirst} % Ensures the first paragraph after a section is indented

\usepackage[letterpaper, left=0.54in, right=0.54in, top=0.65in, bottom=0.94in]{geometry}

\begin{document}
\title{Differentiable High-Order Markov Models \\for Spectrum Prediction}
	\author{
		\IEEEauthorblockN{Vincent Corlay\IEEEauthorrefmark{1},  Tatsuya~Nakazato\IEEEauthorrefmark{2},~Kanako~Yamaguchi\IEEEauthorrefmark{2},~and~Akinori~Nakajima\IEEEauthorrefmark{2}}
			\IEEEauthorblockA{\IEEEauthorrefmark{1}Mitsubishi Electric R$\&$D Centre Europe,~Rennes,~France.~Email:~v.corlay@fr.merce.mee.com}
			\IEEEauthorblockA{\IEEEauthorrefmark{2}Information Technology R$\&$D Center, Mitsubishi Electric Corporation, Ofuna, Kamakura, Japan}
	}
	\maketitle
	\thispagestyle{empty}
%====================================================================================================================================================
\linespread{0.92}
%\tableofcontents
\begin{abstract}
The advent of deep learning and recurrent neural networks revolutionized the field of time-series processing.
Therefore, recent research on spectrum prediction has focused on the use of these tools. 
However,  spectrum prediction, which involves forecasting wireless spectrum availability, is an older field where many ``classical" tools were considered around the 2010s, such as Markov models.
This work revisits high-order Markov models for spectrum prediction in dynamic wireless environments.  
We introduce a framework to address mismatches between sensing length and model order as well as state-space complexity arising with large order. 
Furthermore, we extend this Markov framework by enabling fine-tuning of the probability transition matrix through gradient-based supervised learning, offering a hybrid approach that bridges probabilistic modeling and modern machine learning. 
Simulations on real-world Wi-Fi traffic demonstrate the competitive performance of high-order Markov models compared to deep learning methods, particularly in scenarios with constrained datasets containing outliers.
\end{abstract}

\section{Introduction}
Spectrum prediction is a critical task in wireless communications, particularly for cognitive radio systems and interference management. Accurate spectrum prediction enables more efficient resource allocation, better quality of service, and reduced interference between users. In cognitive radio systems, secondary users (SU) rely on prediction algorithms to anticipate the behavior of primary users (PU) and adapt their communication strategies accordingly. 

As described in the rich survey \cite{Ding2018}, traditional approaches to spectrum prediction include statistical models like autoregressive moving averages (ARMA) and Markov models.
With the rise of deep learning, recurrent neural networks (RNN), including their variants such as long short-term memory (LSTM) networks and gated recurrent units (GRU), have emerged as powerful tools for spectrum prediction \cite{LSTM2017}-\cite{LSTM2024}. These models excel at capturing long-term dependencies and complex temporal relationships in data. 
Some studies focus on one-step predictions \cite{LSTM2017}\cite{LSTM2018}\cite{LSTM2020} and others on multi-step predictions \cite{Shawel2019LSTM}\cite{Gao2021LSTM}\cite{LSTM2023}.
However, neural networks also present significant challenges, such as high computational costs, reliance on extensive labeled datasets, and susceptibility to over-fitting in the presence of noise or outliers. These limitations hinder their applicability in real-time or constrained environments. For instance, \cite{Shawel2019LSTM}\cite{Gao2021LSTM}\cite{LSTM2023} all consider a label-sequence length of the same size as the prediction-sequence length. In Appendix~\ref{ref_appendix}, we show that the prediction error rate of RNN increases if the length differs.
The reader is encouraged to consult the recent surveys \cite{SurveyDL2024_A}\cite{SurveyDL2024_B} for a comprehensive overview of deep learning for spectrum prediction.

In contrast, Markov models (or equivalently Markov chains) offer a simpler yet robust alternative. By focusing on state transitions and leveraging empirical estimation methods, Markov models provide interpretable and computationally efficient solutions. 
As highlighted in \cite[Section III.A]{Ding2018}, these models, often limited to first-order (memory-one) implementations, have been extensively considered for spectrum prediction (or to generate synthetic traffic) before the massive usage of deep learning, see also \cite{HMM2016}\cite{LSTM2021}\cite{LSTM2021B}. 
However, their inability to capture longer-term dependencies restricts their performance in complex traffic scenarios. 
On the other hand, \cite{Chen2010} considers higher-order Markov models to benefit from several past sensing observations to improve the prediction accuracy. 
The order refers to the memory of the model. 
Nevertheless, and surprisingly, research using higher-order Markov models for spectrum prediction remains limited.  
For instance, no paper discusses the case of a sensing length being shorter than the model order\footnote{This is to be differentiated from standard hidden Markov models where noisy versions of the states are observed, as in \cite{Chen2010}.}. 

\textbf{Contributions.} This paper focuses on high-order Markov models for spectrum prediction. We address the following aspects:
\begin{itemize}
\item The need to have a model order greater than the sensing length.% and model order mismatch. 
\item The state-space choice, to handle the exponential growth in the number of composite states associated with higher orders.
%: Strategies are introduced to encode incomplete observations into probabilistic state distributions, ensuring robust predictions.
\item Differentiability: A novel fine-tuning approach using gradient-based supervised learning allows high-order Markov models to adapt to model-choice inconsistencies. %, bridging the gap between probabilistic and neural network-based methods.
\end{itemize}
Through simulations on real-world Wi-Fi traffic data, we demonstrate the competitiveness of high-order Markov models against deep learning methods.
This highlights the potential of high-order Markov models in resource-constrained scenarios.% with constrained datasets containing outliers. %as well as resource-constrained environments. 
%This work not only highlights the potential of high-order Markov models but also emphasizes their practicality for real-time and resource-constrained environments.

The datasets used to conduct the simulations are publicly available at this \href{https://github.com/corlay-MERCE/Datasets-Paper-Differentiable-High-Order-Markov-Models-for-Spectrum-Prediction}{\underline{link}}\footnote{https://github.com/corlay-MERCE/Datasets-Paper-Differentiable-High-Order-Markov-Models-for-Spectrum-Prediction}.

\section{Wi-Fi traffic measurements}

\begin{figure}[h]
	\centering
	\includegraphics[scale=0.62]{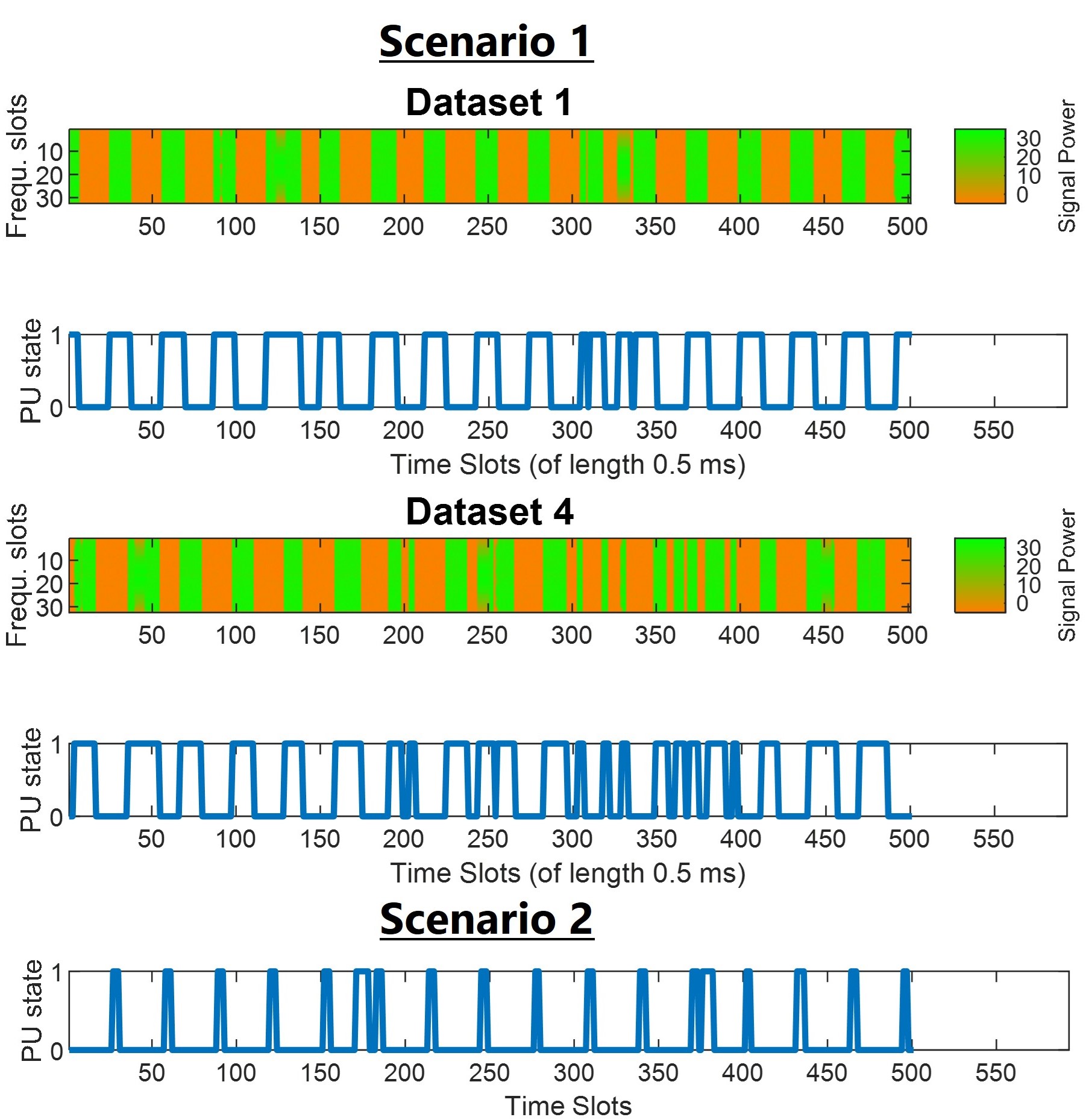}

	\caption{Examples of measured energy levels (green and orange colors) and corresponding PU states (in blue).}
\vspace{-3mm}
	\label{fig:energy_levels}
\end{figure}

We performed measurements of a PU using the  Wi-Fi 802.11ax protocol \cite{WiFiref}.
One single channel of 20 MHz is considered. 
Measurements were collected for two scenarios.
\begin{itemize}
\item Scenario 1: Number of transmitters: 1, number of receivers: 4 , required transmission rate per Rx: 20Mbps, time-slot length: 0.5ms. 
\item Scenario 2: Same as 1 but with only 1 receiver. The number of receivers affects the activation probability of the transmitter.
\end{itemize}
Each scenario contains 5 datasets, obtained at different measurement times. The results of the measurements are energy levels on frequency slots of 0.2 MHz, see Figure~\ref{fig:energy_levels}.

Let us define the PU state at one time slot $t$ as $q_t \in \{0,1 \}$: $q_t=1$ is the ``active" state (transmission) and $q_t=0$ is the ``not active" state (idle). 
With the considered data, the PU state is fully observable: no error occur in identifying the states when using a thresholding approach on the energy levels.
The resulting PU states over time are also shown in Figure~\ref{fig:energy_levels}. 

We observe a quasi-periodic blockwise nature of the traffic, with ``outliers", where the system alternates between transmission periods and idle periods. The overall activation probability is 50$\%$ for scenario 1 and 15$\%$ for scenario 2. To maintain efficiency on all categories of traffic patterns, the periodic information is not used as prior knowledge by the learning algorithms.  In other words, we do not consider Bayesian or parametric estimation approaches.

Note that for scenario 1, dataset 4 presents more outliers than datasets 1,2,3,5 (despite having the same settings). It is therefore an interesting dataset to test the generalization properties of the considered models.

%\begin{figure}[h]
%	\centering
%	\includegraphics[scale=0.45]{traffic_pattern.png}
%	\caption{Traffic pattern obtained with scenario 1 and scenario 2.}
%	\label{fig:traffic_pattern}
%\end{figure}

\section{Sensing model}
\label{sec_model}

We consider a system where the sensing period is of length $M$ time slots. For instance, $M=20$ represents a sensing length of 10 ms. 
The distance between the last sensing slot and the prediction slot is referred to as the prediction horizon $T$. 
%This system is illustrated in Figure~\ref{fig:frame_model}.

In this work, the sensing length $M$ is viewed as a system constraint. 
We investigate the prediction performance as a function of $M$ but we do not investigate the best trade-off between the sensing length and the communication length. The reader is invited to consult our recent work \cite{Corlay2024} if interested in this aspect.

The prediction models are trained on data obtained by observing the traffic at different slots than the sensing time.
In a online training case, the amount and form of such data may be limited, see e.g., the discussion on $T_{train}$ in Section~\ref{sec_rem}.

%\begin{figure}[t]
%	\centering
%	\includegraphics[scale=0.55]{frame_model.jpg}
%	\caption{Considered sensing model.}
%	\label{fig:frame_model}
%\end{figure}

\section{Proposed High-Order Markov Model}

High-order Markov models extend the classical Markov model framework by incorporating longer historical contexts into their state definitions. This section details the construction of the proposed high-order Markov model and its integration into spectrum prediction tasks.

\subsection{Composite-state representation}

In traditional first-order Markov models, the next state is determined solely by the current state. While this simplicity makes first-order models computationally efficient, it often leads to suboptimal predictions in scenarios with temporal dependencies. %, such as block-structured or periodic traffic patterns. 
To address this, high-order Markov models use composite states, which encode sequences of $M'$ past states as:
\begin{align}
\mathcal{Q}_i = [q_{t},...,q_{t-M'+1}],
\end{align}
where $M'$ is the memory length, also called order, of the model. %By including longer temporal histories, the model captures richer dynamics of the traffic patterns. 
Then, a high-order Markov model can be represented by a 2D transition probability matrix $P$.
The entry $P_{ij}$ at row $i$ column $j$ represents the transition probability between a composite state $\mathcal{Q}_i$ at time $t$ and a subsequent composite state $\mathcal{Q}_j$ at time $t+1$: $P_{ij}=p(\mathcal{Q}_j|\mathcal{Q}_i)$. The only non-zero entries in $P$ are of the form $\mathcal{Q}_j =[q_{t+1},...,q_{t-M'+2}]$ and $\mathcal{Q}_i =[q_{t},...,q_{t-M'+1}]$. 
%The size of the transition matrix grows as $2^{M'}\times 2^{M'}$. % which can become computationally intractable for large $M'$. To mitigate this, we explore techniques for state-space reduction,

Each composite state $Q_i$ is associated to an index $i$. The $i$-th value of the composite-state probability vector $s_t$ gives the probability of being in the composite state $Q_i$ at time $t$:
\begin{align}
s_t[i] = \text{Probability of being in composite state } \mathcal{Q}_i \text{ at time $t$}.
\end{align}
The size of $s_t$ is equal to the number of composite states.
Note that this modeling is different (but equivalent) from \cite{Chen2010}, where a $M'-$dimensional tensor is used rather than a large 2D matrix.

\subsection{Model training and inference}

Training the transition matrix $P$ involves counting occurrences of state transitions in the training data: 
\begin{align}
p(\mathcal{Q}_j|\mathcal{Q}_i) = \frac{\text{Number of transitions from} \ \mathcal{Q}_i \text{ to } \mathcal{Q}_j}{\text{Total occurrences of } \mathcal{Q}_i}.
\end{align}
We refer to this training algorithm as empirical estimation.
Algorithms such as Baum-Welch are not required as the states are fully observable. Note that empirical estimation requires labeled data only for the next immediate time step (as opposed to using full sequences as labels, see Section~\ref{sec_benchmarkNN}). 
%Empirical estimation allows for rapid updates, making it ideal for online training scenarios where models need to adapt dynamically to new data.

The inference is then performed as follows. 
First, the sensing signal of length$M$ at time $t$ is encoded in the corresponding composite-state probability vector $s_{t}$.
%\footnote{This is different from the standard HMM where the $M=M'$ but where one sees a noisy version of the state as e.g., in \cite{Chen2010}}.  
If $M'>M$, the sensing signal may correspond to several composite states which is reflected in the encoded vector $s_t$:
If there are $N$ composite-state candidates, then the vector $s_t$ contains the probability value $1/N$ at the index of each candidate.
 If $M'=M$, each sensing signal corresponds to a unique composite state $\mathcal{Q}_i$ and $s_t$ contains a 1 at position $i$ and 0 elsewhere.
Then, the composite-state prediction at time $t+T$ is computed as 
\begin{align}
s_{t+T}=s_t \times P^T.
\end{align}
Finally, the probability to have a PU active state $q_{t+T}=1$ at time $t+T$ is obtained by summing the values of $s_{t+T}$ corresponding to any composite state of the form $\mathcal{Q}_i =[q_{t+T}=1,...]$.

\subsection{Toy example to justify $M'>M$ in case of periodic traffic with long block size}
\label{Sec_toy_example}
The following example illustrates why one should sometimes choose $M'\geq M$, especially if the block size of a periodic traffic is greater than the sensing length $M$.
Consider a PU traffic pattern with deterministic blocks of size 3:  $ […111 0 0 0 1 1 1 0 0 0 1 1 1 0 0 0  … ]$. First, with $M=M'=3$, there are 6 possible distinct sensed vectors. 
Error-free spectrum prediction is always possible. Then, consider a sensing length $M=2$ and a model memory $M'=3$. Assume that the vector $[11]$ is sensed at time $t$. It could correspond to two composite states:
\begin{itemize}
\item If the (unseen) composite state is $[0 11]$, the prediction at $t+1$ should be 1 and the prediction at $t+2$ should be 0.
\item If the (unseen) composite state is $[1 11]$, the prediction at $t+1$ should be 0 and the prediction at $t+2$ should be 0.
\end{itemize}
Hence, there is uncertainty for the prediction at time $t+1$ but not for time $t+2$: it is always 0.
If the model memory is $M'=2$, a good prediction at $t+2$ is not possible.

\begin{figure}[h]
	\centering
	\vspace{-3mm}
	\includegraphics[scale=0.65]{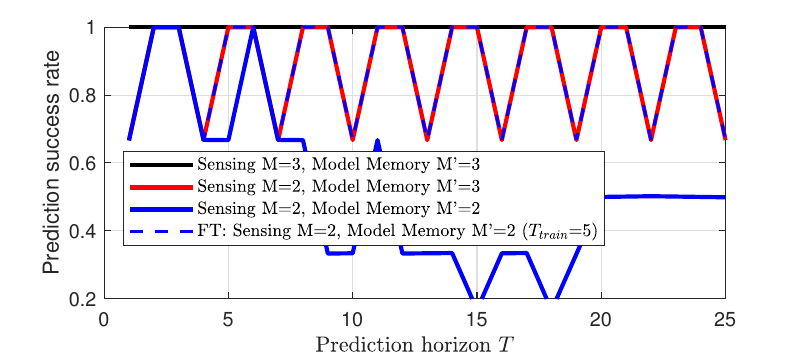}
	\caption{Simulation results for the toy example using deterministic PU traffic with blocks of size 3, to illustrate the need of $M'\geq M$.}
	\label{fig:simu_res_toy_example}
\end{figure}

Figure~\ref{fig:simu_res_toy_example} illustrates the prediction performance, as a function of the prediction horizon $T$, considering these cases. 
Each point for a given value of $T$ represents the average performance obtained with different sensing signals. 
It confirms that a model memory $M'=3$ is required even if $M=2$.
As expected, we also observe that when $M<M'$ (case of long block size and short sensing length), there are remaining unavoidable ambiguities at some time steps due to the non-uniqueness of the composite states corresponding to a given sensed vector. 
The dashed blue curve ``FT" is discussed in Section~\ref{diff_Markov_chain}.

 %This explains the periodic performance drops in Figure~\ref{fig:res_scen1} and \ref{fig:simu_res_toy_example}.

\subsection{Choice of the state space}

With a ``full-state" approach, considering all possible binary vectors, the number of composite states grows as $2^{M'}$. 
The considered blockwise traffic requires large $M'$, up to 30. The full-state approach is therefore not tractable. 
Sub-sampling the datasets, to reduce the required order, significantly degrades the performance of all models (including neural networks) and is therefore not explored.
%It is necessary to simplify the state space. 

Considering the blockwise nature of the traffic, a first solution is to choose the composite states as the number of last time slots in the same active or inactive PU state in the sensing vector. 
%This is the one used in the toy example of the previous subsection. 
This yields only  $2 \times M'$ states and we therefore call it ``simple Markov". 
However, this state-space choice is likely to be penalized by outlier sequences such as [...11111000100...]. %(the one mentioned is Section~\ref{sec_preli_comments}).

Alternatively, the composite-state space can be discovered in the learning process as follows. 
Before learning the transition matrix $P$,  a table with the distinct sensing signals encountered in the training dataset is created and a composite-state index is associated to each signal. 
The probability transition matrix is then computed via empirical estimations on these found composite states. 
During the inference process, if a composite state is not in the table, the ones having the closest Hamming distances are considered.
Since the data has a periodic pattern with a limited number of outliers, the number of composite states should remain limited. 
We refer to this approach as ``smart state Markov".

Note that during the learning process, a limit to the maximum number of composite states in the table can be set. 
Then, the challenge amounts to choosing the best representatives. This has similarities with dictionary-learning problems \cite{Ivana2011} and it is an interesting line of research that is left for future work.

%Finally, subsampling could also be considered to reduce required $M'$ and make the full state approach tractable. 
%Results provided in Appendix show that it leads to performance degradation even with full supervised learning (Figure~\ref{fig:res_scen1}).

\section{Differentiable Markov Models}

The proposed differentiable Markov models extend the classical Markov modeling framework by introducing the ability to fine-tune parameters, such as the transition probability matrix, through gradient-based optimization. 
This enhancement enables to compensate for potential mismatches in initial model configurations.
As explained in the previous subsection, the mismatch can be either due to a wrong model-memory choice $M'$ and/or a wrong state-space choice.

\subsection{Markov model as an embedding approach}
\begin{figure}[h]
	\centering
	\includegraphics[scale=0.42]{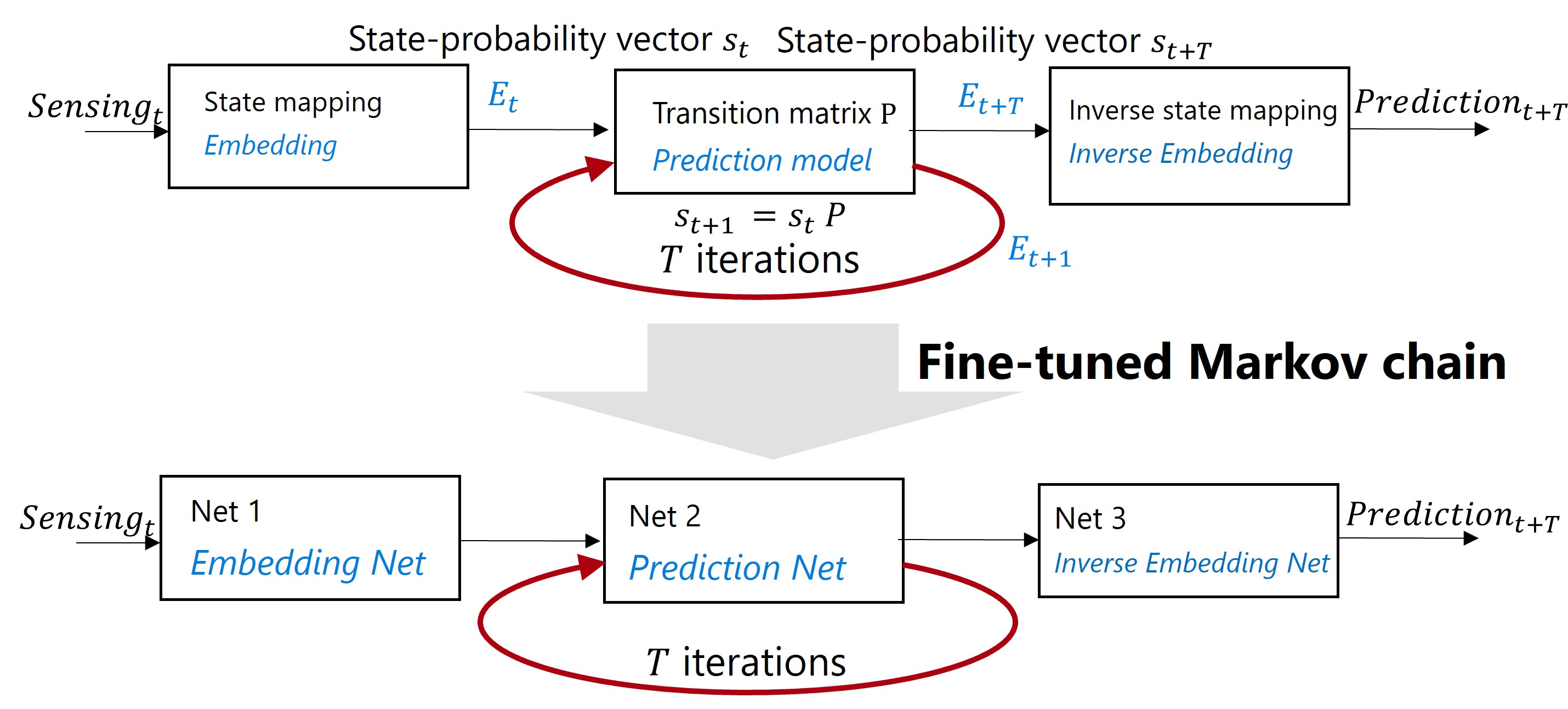}
\vspace{1mm}
	\caption{Representation of a Markov model as an embedding approach. Top: standard Markov, bottom: differentiable Markov. }
\vspace{2mm}
	\label{fig:Markov_chain_embed}
\end{figure}

Markov models can be interpreted as an embedding approach, as illustrated in Figure~\ref{fig:Markov_chain_embed}. 
Consequently, each block or some of the blocks of this model could  be replaced by a neural network.

This approach aligns with a broader trend of transforming existing algorithms into differentiable forms, enabling their optimization and enhancement through supervised training techniques.
Examples include deep unfolding methods (e.g., for the belief-propagation algorithm) \cite{BalatsoukasStimming2019}\cite{Corlay2018} or improving the MUSIC algorithm in case of hardware impairments \cite{Chatelier2024}. 

\subsection{Fine-tuning the probability transition matrix $P$}
\label{diff_Markov_chain}

The model proposed in Figure~\ref{fig:Markov_chain_embed} enables fine-tuning of the probability transition matrix $P$ via supervised learning. 
As a first use of this approach, we consider only the replacement of $P$ by Net2. 
Net2 is therefore a matrix initialized as the transition matrix $P$ learned via empirical estimation. 
It is then fine-tuned in a supervised-training manner. %using the constrained dataset where $T\in [1, \  T_{train}]$.

The training dataset for supervised training is composed of composite-state probability vectors obtained by implementing a sliding window on the training traffic trace.
For each time step $t$, the last $M'$ PU states are used to compute the composite state $\mathcal{Q}_i$, which is encoded in the corresponding composite-state probability vector~$s_t$.
The corresponding labels are the following $T_{train}$ composite-state probability vectors of the traffic trace $s_{t+1}$,...,$s_{T_{train}}$.
As a result, the training dataset is composed of many pairs 
\begin{align}
( \text{input: }s_t, \text{ labels: } s_{t+1},...,s_{t+T_{train}} ).
\end{align}
Algorithm~\ref{alg:differentiable_markov} summarizes the fine-tuning training process. We refer to this approach as ``fine-tuned (FT) Markov".
%For this fine tuning step, the composite state index are kept unchanged and the labels are the composite state probability vectors of the considered state space.

\begin{algorithm}
\caption{Training process for fine-tuned Markov}
\label{alg:differentiable_markov}
\begin{algorithmic}[1]
\REQUIRE Transition probability matrix $P$ learned via empirical estimation. \\ Training dataset $\{( \text{input: }s_t, \text{ labels: } s_{t+1},...,s_{t+T_{train}} ) \}$. %, learning rate $\eta$, maximum epochs $E$, loss function $\mathcal{L}$.

\STATE \textbf{Initialization:}
\FOR{epoch $e = 1$ to $E$}
    \STATE \textbf{Forward Pass:}
    \FOR{each input sample $s_t$}
	\FOR{\( n = 0 \) to \( T_{\text{train}}-1 \)}
           	 \STATE Compute  $\hat{s}_{t+n+1} = \hat{s}_{t+n} P $ (where $\hat{s}_t=s_t$).
        	\ENDFOR
        %\STATE Compute predicted probability vector $\hat{s}_j = P^T s_i$.
    \ENDFOR

    \STATE \textbf{Loss Calculation:}
    \STATE Compute the loss as \\ $\mathcal{L} = \sum_t \mathcal{L}( [\hat{s}_{t+1},...,\hat{s}_{t+T_{train}}], [s_{t+1},...,s_{t+T_{train} } ] )$.
    \STATE \textbf{Update the matrix:}
    \STATE For all for all entries of $P$, compute gradients $\frac{\partial \mathcal{L}}{\partial P_{ij}}$. Update $P_{ij}$ using gradient descent :$P_{ij} \leftarrow P_{ij} - \eta \frac{\partial \mathcal{L}}{\partial P_{ij}}$, where $\eta$ is the learning rate (Adam optimization can also be used).
\ENDFOR
\RETURN Fine-tuned probability transition matrix $P$.
\end{algorithmic}
\end{algorithm}
 \vspace{-2mm}

One potential drawback of this approach is the risk of losing generalization capabilities beyond the training horizon $T_{train}$.
Numerical simulations are therefore necessary to verify if $(i)$ improvement is observed for $T\in [1, \ T_{train}]$ $(ii)$ generalization beyond $T_{train}$ is effective.

Regarding the training complexity, it is very fast. Only a couple of iterations are required for the fine-tuning step to reach a quasi-constant loss value.

\section{Simulation results}

\subsection{Benchmark (neural network)}
\label{sec_benchmarkNN}
As benchmark, we consider a standard feed-forward neural network trained in supervised manner for a prediction horizon $T\in [1, \ T_{train}]$, with $T_{train}=T_{max}$ where $T_{max}$ is the largest desired prediction horizon. 
We also tested LSTM-based RNN. However, they did not demonstrate any significant advantage over standard feed-forward neural networks, see Appendix~\ref{ref_appendix}.

The size of the input of the neural network is $M$ and the size of the output is $T_{train}$. 
It has three fully-connected hidden layers of size 80 with ReLU activation functions. The Adam optimization algorithm with a learning rate of 0.001 is used.
The training dataset is composed of many couples 
\begin{align}
(\text{input: sensing vector, label: PU states for } T \in [1, \ T_{train}] ),
\end{align}
obtained by implementing a sliding window on the training traffic trace.  
The neural network is trained to minimize the squared error between its outputs and the labels.

\subsection{Remarks on the training dataset}
\label{sec_rem}
The feed-forward neural network cannot output predictions for values $T$ out of the training range $[1, \ T_{train}]$.
Moreover, we report in Appendix~\ref{ref_appendix} simulations showing that RNN, which can output predictions beyond $T_{train}$ in a closed-loop mode, are not robust to reduced-size labeling sequences (small value of $T_{train}$).

On the other hand, with empirical estimation of the probability transition matrix $P$, we have $T_{train}=1$. 
This offers flexibility in cases of constrained datasets where full sequences to be predicted are not available as labels. 
This can be considered as a semi-supervised learning mode.
The learning algorithm (empirical estimation) is also significantly faster than back-propagation, which is an advantage for online training with dynamic PU behavior.
%With a large enough memory, the prediction performance remains high even with large $T$, at around 90 $\%$. 
%The performance with other test datasets is similar (except for dataset 4 of scenario 1) and therefore not displayed.

\subsection{Simulation results of high-order Markov models}

The results are presented as a function of the prediction horizon $T$. 
Each point for a given value of $T$ represents the average performance obtained with different sensing signals. 
Unless otherwise specified, a different dataset is used for training and inference (but from the same scenario).

\begin{figure}[t]
	\centering
         \vspace{-4mm}
	\includegraphics[scale=0.62]{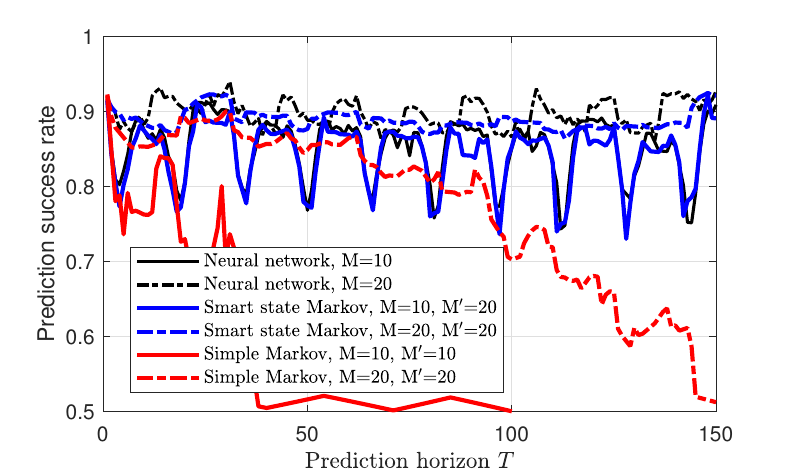}
	\caption{Prediction performance of the neural network and Markov models with different state-space choices and different model memory $M'$. Scenario 1: training on dataset 1 testing on dataset 3.}
\vspace{-1mm}
	\label{fig:simu_res_Markov}
\end{figure}

The prediction performance with scenario 1 is presented in Figure~\ref{fig:simu_res_Markov}. The performance with the other test datasets 2 and 5 is similar and therefore not displayed.
%On the one hand, with a good model choice, the Markov approaches offer the same performance as the neural network.
On the one hand, we observe that a wrong model-memory choice and/or state-space choice has a high negative impact on the performance. 
%As expected,  the ``simple Markov" state-space model is not appropriate. 
On the other hand, smart-state Markov exhibits performance similar to the neural network.
The periodic performance drops with $M=10$ (observed both with smart-state Markov and the neural network) are due to the composite-state uncertainty explained in Section~\ref{Sec_toy_example}. % with full supervised training. 

\begin{figure}[t]
	\centering
	\includegraphics[scale=0.6]{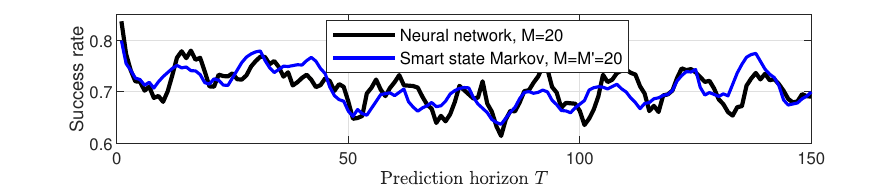}
	\caption{Prediction performance of the neural network and smart-state Markov. Scenario~1: training on dataset 1 testing on dataset 4.}
\vspace{-1mm}
	\label{fig:simu_res_Markov_2}
\end{figure}
\begin{figure}[t]
	\centering
	\includegraphics[scale=0.6]{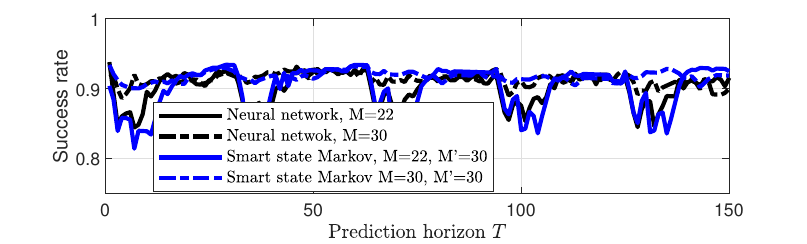}
	\caption{Prediction performance of the neural network and smart-state Markov. Scenario~2: training on dataset 1 testing on dataset 2.}
\vspace{-2mm}
	\label{fig:simu_res_Markov_3}
\end{figure}

Figure~\ref{fig:simu_res_Markov_2} reports the performance when testing on the more random dataset 4 of scenario 1 (see Figure~\ref{fig:energy_levels}) and Figure~\ref{fig:simu_res_Markov_3} reports the performance with scenario 2.
Again, the smart-state Markov approach achieves similar performance as the neural network. We note that, as expected, the performance of both models is degraded with dataset 4 of scenario 1.

The smart-state Markov approach is therefore competitive with a feed-forward neural network while being more flexible on the training aspect.

\subsubsection{Number of composite states with smart-state Markov} When training on dataset 1 (scenario 1) with $M=M'=20$, the number of states added to the table is $L=120$. The average prediction success rate on dataset 3 for $T\in [1, \ 150]$ is 88$\%$ (dashed blue curve in Figure~\ref{fig:simu_res_Markov}). If the maximum number of states is set to $L=40$ (added on a ``first come first added" basis), the performance remains as high as $87\%$, but for $L=30$ it suddenly drops to 53$\%$. Again, the optimization of the state space is left for future work. 
However, potential mistakes on the value of $L$ can be corrected via the fine-tuning approach.

\subsection{Simulation results of the FT Markov model}

The simulation result with the toy example is provided in Figure~\ref{fig:simu_res_toy_example}: 
with $M=M'=2$, fine-tuning enables to achieve the same performance as $M=2$, $M'=3$. 
In other words, it successfully corrects the wrong model-memory choice.

\begin{figure}[t]
	\centering
	\includegraphics[scale=0.62]{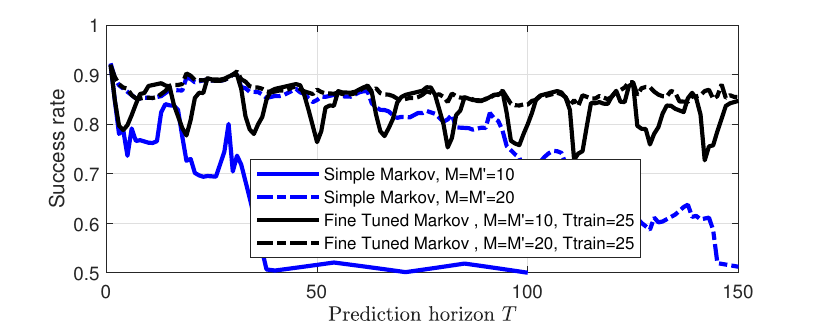}
	\caption{Simulation results of FT Markov. Scenario 1: training on dataset 1 and testing on dataset 3. Case of bad initial model choice: ``Simple Markov".}
\vspace{-1mm}
	\label{fig:simu_res_Markov_FT_1}
\end{figure}

\begin{figure}[t]
	\centering
	\includegraphics[scale=0.62]{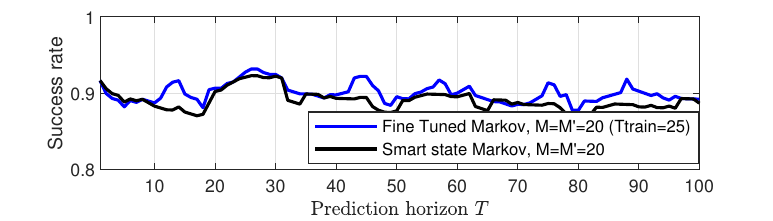}
	\caption{Simulation results of FT Markov. Scenario 1: training on dataset 1 and testing on dataset 3. Case of good initial model choice: smart-state Markov.}
	\label{fig:simu_res_Markov_FT_2}
\end{figure}

The results with the realistic data are provided in Figure~\ref{fig:simu_res_Markov_FT_1} (case of bad initial model choice) and Figure~\ref{fig:simu_res_Markov_FT_2} (case of good initial model choice). 
Similarly, the approach successfully corrects wrong model state-space choice and offers good generalization beyond $T_{train}$.
If the choice of the model is already correct the method still improves marginally the performance.


\subsection{Simulation results on the generalization aspect}

\begin{figure}[t]
	\centering
	\includegraphics[scale=0.6]{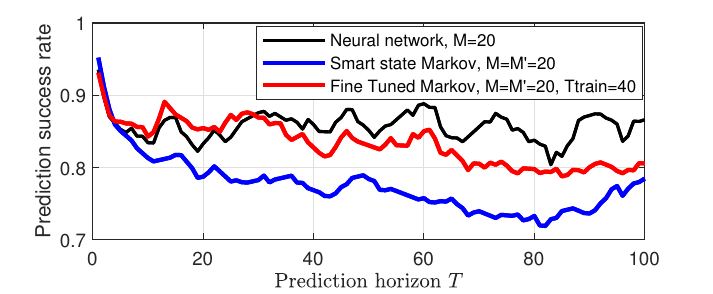}
	\caption{Simulation results when training on dataset 4 (scenario 1). Testing on dataset~4.}
	\label{fig:simu_res_Markov_FT_1_dataset4_1}
\end{figure}

\begin{figure}[t]
	\centering
	\includegraphics[scale=0.58]{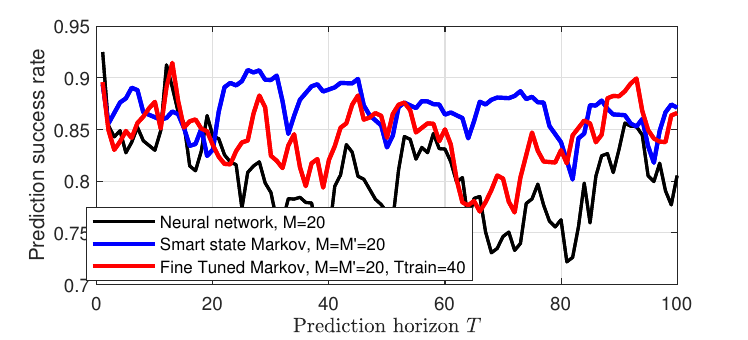}
	\caption{Simulation results when training on dataset 4 (scenario 1). Testing on dataset~3.}
	\label{fig:simu_res_Markov_FT_1_dataset4_2}
\end{figure}

We use dataset 4 of scenario 1 to investigate the generalization performance of the models. The neural network, smart-state Markov, and FT Markov models are considered.
The results are provided in Figure~\ref{fig:simu_res_Markov_FT_1_dataset4_1} and Figure~\ref{fig:simu_res_Markov_FT_1_dataset4_2}.

The neural network is the method that best specializes to dataset 4, but has poor generalization performance on dataset 3. This is a typical over-fitting behavior. On the other hand, the Markov approach does not well specialize on dataset 4 but manages to get the key pattern of scenario 1 as it generalizes well to dataset 3. 
Finally, FT Markov offers the best trade-off between specialization and generalization. Note that this trade-off can be controlled as a function of $T_{train}$, where a greater value yields better specialization and a smaller value yields better generalization.

\section{Conclusion}

In this paper, we investigated high-order Markov models for spectrum prediction. These models allow for accurate prediction of periodic traffic patterns in cognitive radio systems, providing a promising alternative to deep learning-based models, especially in scenarios with limited or constrained datasets.
%Additionally, the proposed model addresses critical issues such as partial observability and state-space explosion, offering scalable and efficient solutions. Our method shows significant promise for spectrum prediction in practical cognitive radio systems, where accurate, real-time decisions are paramount.

Through the introduction of differentiable Markov models, model fine-tuning via gradient-based optimization is enabled. 
Simulation results demonstrate that bad initial model choices are successfully corrected in the fine-tuning step.

\textbf{Future work.} 
The framework proposed in Figure~\ref{fig:Markov_chain_embed} offers other possibilities than the one evaluated to improve a Markov model.  
For instance, the differentiable approach could be used for more efficient state mapping and demapping.

Moreover, in case of limited number of states to be considered in the smart-state Markov model, the selection of the best representatives as a relevant dictionary is an interesting area of research.

Finally, the proposed algorithms should also be investigated in a more advanced multi-channel situation.
Moreover, we plan to investigate not only the hard 0/1 predictions but also the probability estimates, useful for efficient link adaptation.

%We leave the exploration of these possibilities for future studies.
%This integration enables the transition matrix to be fine-tuned using gradient-based optimization. 
%, offering a hybrid approach that bridges the gap between traditional statistical models and modern neural networks. The experimental results demonstrate that high-order Markov models, particularly with fine-tuning, outperform conventional models and exhibit robustness to noisy data and outliers, making them suitable for real-time spectrum prediction tasks.

%In future work, we plan to extend this approach to multi-channel spectrum prediction, investigate its application to dynamic wireless networks with evolving traffic.

\section{Appendix: Predictions with recurrent neural networks}
\label{ref_appendix}

The aim of this appendix is to provide evidence that RNN are not robust to small values of $T_{train}$ (size of the labeling sequence), as claimed in Section~\ref{sec_benchmarkNN}.

\subsection{Background on RNN and training protocol}

At each time slot, a RNN takes one sample as input and outputs one sample.
In an open-loop mode, the inputs are data from an observed sequence. 
In a closed-loop mode, the input is the output of the previous iteration.

\begin{figure}[t]
	\centering
	\includegraphics[scale=0.325]{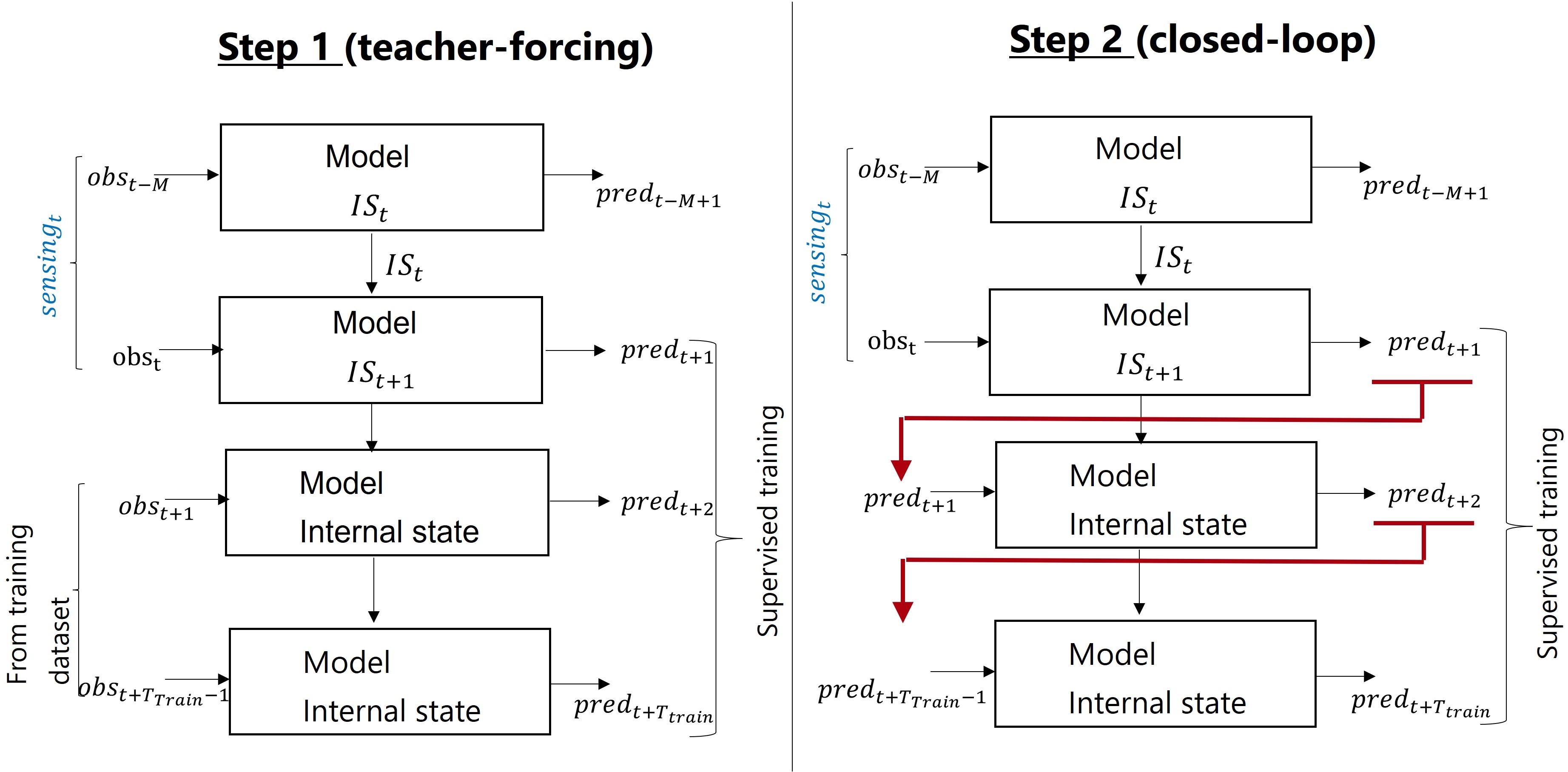}
	\caption{Two-steps training protocol for the RNN. IS stands for internal state.}
	\label{fig:training_protoc_RNN}
\end{figure}

For the considered task, the open-loop mode is thus used for the first $M$ iterations of the model, where the sensed samples are provided.
The outputs of these $M-1$ first iterations are not used. % as the true sensing values are available.
Starting from the $M+1$ iteration, the closed-loop mode is used (see the second column of Figure~\ref{fig:training_protoc_RNN}).
As a result, predictions for any value $T$ can be generated by using $T-1$ closed-loop iterations.
The described structure is equivalent to the encoder-decoder architecture usually considered for sequence-to-sequence regression problems.
 
%Unlike Markov models, RNN can be classified as ``internal-state space" approach, to be opposed with the embedding approach. 
%The input/output data passed at each iteration contains a low amount of information on the state of the system. 
%This information is in the state of the model, which is updated as each iteration as shown on Figure~\ref{fig:training_protoc_RNN}.

%\subsection{}

The training process is illustrated in Figure~\ref{fig:training_protoc_RNN}. 
We consider the scheduled sampling approach \cite{Bengio2015}.
It contains two steps. In the first step,  only samples for the dataset are used as input (open-loop mode only). 
Supervised learning is performed on the outputs $t+1$ to $t+T_{train}$.
%The goal of this step is that the neural network learns to perform correct predictions and store relevant information in the internal state.
In the second step, the RNN is trained using its own prediction as input from time $t+1$ (iteration $M$), as done during the inference step.

\begin{figure}[t]
	\centering
\vspace{-1mm}
	\includegraphics[scale=0.62]{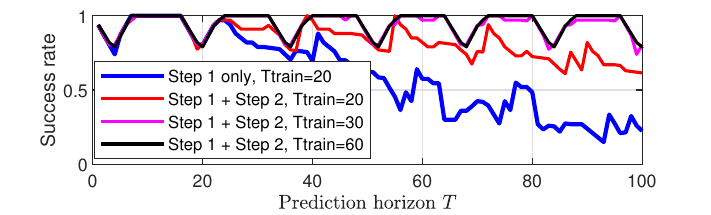}
	\caption{Performance of the RNN as a function of $T_{train}$ with the toy example (block size 16).}
	\label{fig:RNN_perf_toy_example}
\end{figure}

\begin{figure}[t]
	\centering
\vspace{-1mm}
	\includegraphics[scale=0.59]{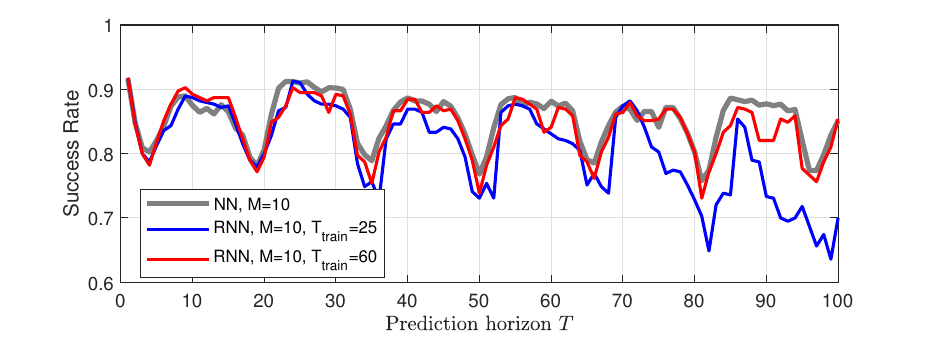}
	\caption{Performance of the RNN as a function of $T_{train}$ with the real data. Scenario 1: training on dataset 1 testing on dataset 3. NN denotes feed-forward neural network.}
	\label{fig:RNN_perf_true_data}
\end{figure}

Note that training such a RNN is more difficult than a standard neural network and less suited to an online implementation: 
It involves several hyper-loops (steps 1 and 2 of Figure~\ref{fig:training_protoc_RNN}) and training in step 2 may be unstable due to the feedback.
There is more variability in the training process than for a standard feed-forward neural network.
The training takes significantly more time than the fine-tuning of the transition matrix of Markov models. 
This was also reported in \cite{LSTM2018} where it is observed that LSTM consumes more than 5 times the training time of feed-forward neural networks.
\cite{LSTM2021B} also reports high complexity of LSTM training and discusses efficient initialization methods to mitigate the problem.

\subsection{Simulation results}
The model architecture of the RNN consists of a sequence input layer, followed by two LSTM layers with 128 units each. One fully connected layers of 128 neurons is used between the two LSTM layers and another one before the output. 

We first consider a toy example with deterministic blocks of size 16 and a sensing length $M=10$ (similar to the values of scenario 1).
Figure~\ref{fig:RNN_perf_toy_example} provides the result.
%If difference between $N$ and $M$ is too large error-propagation phenomenon appears.
We observe that if $T_{train}$ is too small, the performance is degraded.
%Phenomenon attenuated with larger $T_{train}$ but does not disappear.
The RNN struggles to perform correct predictions if trained with labels having a reduced number of pattern periods. %does not manage to retain relevant information from the original sensing vector.
%Struggle with long-term dependencies, which is a known drawback of RNN. 
%Disappointing performance is to be expected with real data.
Figure~\ref{fig:RNN_perf_true_data} provides the performance with the real data. As expected, the performance is disappointing if $T_{train}$ is too small. 

%Note that the performance of the RNN could probably be improved by considering advanced scheduling function \cite{Bengio2015} to switch between step 1 and step 2.
%However, this is not adapted to an online training process which should remain simple.
%. The potential errors in the feedback loop seems difficult for the RNN to handle.

\flushend

\end{document}